\documentclass[prd,twocolumn,nofootinbib,notitlepage,aps,tightenlines,preprintnumbers,amsmath,amssymb,amsfonts,showpacs,superscriptaddress]{revtex4-2}

\RequirePackage[colorlinks=true
,urlcolor=blue
,anchorcolor=blue
,citecolor=blue
,filecolor=blue
,linkcolor=blue
,menucolor=blue
,linktocpage=true
,pdfproducer=medialab
,pdfa=true
]{hyperref}

\usepackage{amsmath,amssymb,amsthm,amsfonts}
\usepackage{graphicx,tabularx}
\usepackage{float}
\usepackage{multirow}  
\usepackage{cancel}
\usepackage{comment}
\usepackage{slashed}
\usepackage{physics}
\usepackage{xspace}

\usepackage{cleveref}
\crefname{section}{Sec.}{Secs.}
\crefname{figure}{Fig.}{Figs.}
\crefname{equation}{Eq.}{Eqs.}
\crefname{appendix}{Appendix}{Appendices}

\RequirePackage[normalem]{ulem}

\newcommand{\bea}{\begin{eqnarray}\begin{aligned}}
\newcommand{\eea}{\end{aligned}\end{eqnarray}}
\newcommand{\gd}{g_d}

\begin{document}
\preprint{FERMILAB-PUB-24-0357-T-V}
	
\title{LANSCE-mQ: Dedicated search for milli/fractionally charged particles at LANL}

\author{Yu-Dai Tsai}
\email{yt444@cornell.edu; yudaitsai@lanl.gov}
\affiliation{Los Alamos National Laboratory (LANL), Los Almos, NM 87545, USA}
\affiliation{Department of Physics and Astronomy,
University of California, Irvine, CA 92697-4575, USA}
\affiliation{Fermi National Accelerator Laboratory (Fermilab), Batavia, IL 60510, USA}

\author{Insung Hwang} 
\affiliation{Department of Physics, Korea University, Seoul, 02841, Korea}

\author{Ryan Schmitz}
\affiliation{Department of Physics, University of California, Santa Barbara, CA 93106, USA}

\author{Matthew Citron}
\affiliation{University of California, Davis, Davis, CA 95616, USA}

\author{\mbox{Kranti Gunthoti}}
\affiliation{Los Alamos National Laboratory (LANL), Los Almos, NM 87545, USA}

\author{Jacob Steenis}
\affiliation{University of California, Davis, Davis, CA 95616, USA}

\author{Hoyong Jeong}
\affiliation{Department of Physics, Korea University, Seoul, 02841, Korea}

\author{Hyunki Moon}
\affiliation{Department of Physics, Korea University, Seoul, 02841, Korea}

\author{Jae Hyeok Yoo}
\affiliation{Department of Physics, Korea University, Seoul, 02841, Korea}

\author{Ming Xiong Liu}
\affiliation{Los Alamos National Laboratory (LANL), Los Almos, NM 87545, USA}

\begin{abstract}

In this paper, we propose an experiment, LANSCE-mQ, aiming to detect fractionally charged and millicharged particles (mCP) using an 800~MeV proton beam fixed target at the Los Alamos Neutron Science Center (LANSCE) facility.
This search can shed new light on numerous fundamental questions, including charge quantization, the predictions of string theories and grand unification theories, the gauge symmetry of the Standard Model, dark sector models, and the tests of cosmic reheating. We propose to install two-layer scintillation detectors made of plastic (such as EJ-200) or CeBr3 to search for mCPs. Dedicated {\textsc{Geant4}\xspace} detector simulations and in situ measurements have been conducted to obtain a preliminary determination of the background rate. The dominant backgrounds are beam-induced neutrons and coincident dark current signals from the photomultiplier tubes, while beam-induced gammas and cosmic muons are subdominant.
We determined that LANSCE-mQ, the dedicated mCP experiment, has the leading mCP sensitivity for mass between $\sim$ 1~MeV to 300~MeV.
\end{abstract}

\maketitle

\tableofcontents

\section{Introduction}

The study of fractionally charged and millicharged particles (mCPs), which carry rational or irrational electric charges, is a strong test of fundamental theories and can help answer the deepest questions in particle physics.  Firstly, it is one of the most powerful tests of the theories proposed to explain empirical charge quantization \cite{Dirac:1931kp, Schwinger:1966nj}, including the grand unification theories (GUTs) \cite{Pati:1973uk}. 
Fractionally charged particles are also one of the most distinctive low-energy signatures of string theory \cite{Wen:1985qj}. Recently, the search for mCPs has also been considered as a way to determine the gauge theory of the Standard Model (SM) of particle physics~\cite{Li:2024nuo,Koren:2024xof}, and as a way to test reheating scenarios in early-Universe cosmology~\cite{Gan:2023jbs}. 
mCPs can also naturally arise from the kinetic mixing between our SM photon and a  beyond the Standard Model (BSM) massless U(1) gauge field~\cite{Holdom:1985ag}, which serves as viable dark matter candidates and can help explain the recently observed cosmic microwave background (CMB) absorption spectrum (see, e.g., \cite{Barkana:2018qrx}). 

MCPs have become an important benchmark model for accelerator experiments~\cite{Davidson:2000hf}. One can directly search for them at LHC (see fractionally charged particle searches at CMS~\cite{CMS:2024eyx}) and CEPC~\cite{Liu:2019ogn}.
Scintillator-based mCP detectors have been developed for deployment at SLAC \cite{Prinz:1998ua}, the LHC (as milliQan~\cite{milliQan:2021lne, Ball:2016zrp, Ball:2020dnx} and FORMOSA~\cite{Foroughi-Abari:2020qar}), Fermilab~\cite{Kelly:2018brz}, and J-PARC \cite{Kim:2021eix}. 
Neutrino experiments such as DUNE, MiniBooNE, and SBND~\cite{Magill:2018tbb} (as well as ArgoNeuT~\cite{Acciarri:2019jly}, FLArE~\cite{Kling:2022ykt}, and SENSEI~\cite{Crisler:2018gci}) also present opportunities to search for mCPs. Other accelerator probes and detector technologies like those of NA64~\cite{Gninenko:2018ter}, LDMX~\cite{Berlin:2018bsc, Akesson:2018vlm}, SENSEI~\cite{Crisler:2018gci}, OSCURA~\cite{Oscura:2023qch}, and SHiP~\cite{Magill:2018tbb} can similarly further the current mCP explorations, along with numerous studies of millicharged dark matter with astrophysical and cosmological observations~\cite{Vogel:2013raa,Vinyoles:2015khy,Chang:2018rso, Liu:2019knx,Li:2020wyl, Harnik:2020ugb} and with direct-detection experiments~\cite{Deniz:2009mu, Ge:2017mcq, Singh:2018von, Banik:2021kpz, Banik:2021kpz, LZ:2022lsv,SuperCDMS:2022zmd, SuperCDMS:2023sql, LZ:2023poo, Boddy:2024vgt, CONNIE:2024off}.

In this paper, we propose an experiment, LANSCE-mQ, aiming to detect the mCPs using an 800 MeV proton beam fixed target at LANSCE's Lujan Center~\cite{2006NIMPA.562..910L}. Leveraging existing facilities as well as ever-maturing detector technologies, LANSCE-mQ can provide one of the strongest searches for low-mass mCPs, shedding new light on fundamental theories, reheating cosmology, and dark-sector models, and position LANL at the forefront of this crucial global effort.

\section{Millicharged Particles}

Here, we discuss models of mCP, $\chi$, which is assumed to be a fermion throughout this paper. Scalar mCPs would yield similar results in accelerator studies.

\subsection{Pure mCP}

The so-called ``pure" mCP, as discussed in \cite{Gan:2023jbs}, can originate from string theory~\cite{Wen:1985qj,Shiu:2013wxa} (predicting fractionally charged particles) or from BSM theories that predict violation of charge quantization. For the ``pure" mCP, a dark photon is not needed, and the Lagrangian of the mCP is:
\bea \label{eq:minimal MCP} \label{eq:minimal mcp lagrangian}
\mathcal{L} \supset i \overline{\chi} (\slashed{\partial} - i g' \varepsilon_\chi \slashed{B} + m_\chi) \chi - \frac{1}{4} B_{\mu \nu} B^{\mu \nu}.
\eea
Here, $B$ is the SM Hypercharge $U(1)_Y$ gauge boson, and $g'=e/\cos\theta_w$ is the gauge coupling of $U(1)_Y$ and $e$ is the electron electric charge. $B = \cos \theta_w A - \sin \theta_w Z$ after the electroweak symmetry breaking. $\chi$ is coupled to both the SM photon $A$ and the $Z$ boson accordingly and becomes a particle with an effective small electric charge $Q_\chi$, and $\varepsilon_\chi\equiv Q_\chi/e$.

\subsection{Effective dark photon mCP}

mCPs can also be generated effectively through a massless dark photon theory. 
The theory includes an additional dark gauge symmetry $U(1)_d$ with the gauge field, dark photon $A'$, coupled to a new particle $\chi$ with a coupling $g_d$, and we define $\alpha_d \equiv g_d^2/4 \pi$.

One then introduces a kinetic mixing between $A'$ and SM hypercharge $U(1)_Y$ gauge field $B$, that can originate from integrating out heavy fermions charged under both $U(1)_Y$ and $U(1)_d$. We have
\bea
\label{eq:mcp_KM_L}
\mathcal{L} & \supset i \overline{\chi}(\slashed{\partial} - i \gd \slashed{A}' + m_\chi) \chi \\
& \quad - 
\frac{1}{4} B_{\mu \nu} B^{\mu \nu} - \frac{1}{4} A'_{\mu \nu}A'^{\mu \nu} + \frac{\epsilon}{2 \cos \theta_w} B_{\mu\nu} A'^{\mu \nu}.
\eea

$B_{\mu \nu}$ is the usual field strength of $U(1)_Y$, and $\epsilon$ is the kinetic mixing between $U(1)_Y$ and $U(1)_d$, with $A'_{\mu \nu} = \partial_\mu A'_\nu - \partial_\nu A'_\mu$. 
When $A'$ is massless, one can choose a basis diagonalizing the kinetic terms (assuming $\epsilon\ll 1$ and up to $O(\epsilon^2)$) under the field redefinition $A' \rightarrow A' + \frac{\epsilon}{\cos \theta_w} B$. 
The Lagrangian becomes 
\bea
\label{eq:chi_B_coupling}
    \mathcal{L} \supset g' \varepsilon_{\chi}\overline{\chi} \gamma^\mu \chi  B_{\mu}, \,\, \text{where $\varepsilon_{\chi} = \frac{\epsilon g_d}{e}$ },
\eea
and again $g'=e/\cos\theta_w$.
Following the aftermath of electroweak symmetry breaking, $B = \cos \theta_w A - \sin \theta_w Z$, $\chi$ becomes an mCP with an effective fraction of electric charge $\varepsilon_\chi\equiv Q_\chi/e$.

\section{The LANSCE facility and LANSCE-mQ detector}
\label{sec:facility}
The LANSCE facility~\cite{2006NIMPA.562..910L} uses a 1 Mega Watt linear proton accelerator that produces an 800~MeV proton beam. At a rate of 120 pulses per second, 800 MeV proton beam pulses are delivered to different neutron spallation targets. Each pulse is a 625~\textmu s long macropulse separated by 8.3~ms. A macropulse contains micropulses with a five-nanosecond gap between them. Twenty macropulses per second are directed toward the Lujan Neutron Scattering Center, where it impinges on a 1L tungsten spallation target. A beam current monitor (BCM) detects these pulses. The BCM provides a copy of transistor-transistor logic (TTL) for all flight paths in the Lujan center. Using BCM, we can measure each pulse's width, height, and total charge individually. 

The 800~MeV proton beam is directed toward the 1L target after passing through the proton storage ring (PSR). The beam from the PSR is high-intensity and has a triangular pulse shape with approximately 290 ns at the base, with a 1 volt amplitude. The 1L target receives 20 pulses every second, each containing $3.1\times10^{13}$ protons and spaced 50~ms apart. 
The Lujan proton target generates a significant amount of neutral pions, which have a very short lifespan of $8.5\times 10^{-8}$ ns and decay while in flight, rather than being halted by nuclear interactions. 
The total number of pions, $N_{\pi^0}$, is directly correlated to the quantity of protons on the target (POT), at the Lujan target, $N_{\pi^0} = 0.115 \times  \rm POT$ \cite{CCM:2021leg}. While the distribution of pions is mostly isotropic, there is a slight increase in the forward direction. Neutral pions mostly decay into two photons, and one of the photons can decay into two mCPs ($\pi^0$ $\rightarrow$ $\gamma$ $\chi$ $\overline{\chi}$). In our analysis, we also take into account $\eta$ meson and the three-body decays that produce the mCPs.

{\bf Detector sites:} The LANSCE-mQ operation does not require a specific flight path and will not disrupt any ongoing nuclear physics and material science programs at Lujan Center. The LANSCE-mQ will be conducted with the detector installed at a distance from the Lujan target of around 6 meters (ER1) and 35 meters (ER2), and operate over 3 years with a corresponding POT of $5.9 \times 10^{22}$.

The LANSCE-mQ detector will consist of $n$ layers, nominally 2 but can range from 1 to 3, each containing 100 plastic scintillator bars measuring 5~cm $\times$ 5~cm $\times$ 150~cm. 
In total, the detector will have 200 bars. Each of these bars will be optically coupled to a high-gain photomultiplier tube (PMT). Segmentation of the volume is beneficial in reducing background interference caused by the PMT dark currents and cosmic muons, as well as utilizing the directionality of incident mCPs to suppress non-pointing particles further.

\subsection{Alternative designs}

The nominal design of LANSCE-mQ would be with double-layer plastic scintillators. To increase the signature rate, one can also require the scintillation in just one layer, but better shielding and background reduction may be needed for a single-layer design.

Here, we list several alternative designs. The goal is to keep the dark rate as low as possible without sacrificing the photon yield.
Also, a smaller volume of detector would allow a smaller neutron background rate, which is one of the dominating backgrounds for LANSCE-mQ.

\begin{itemize}
    \item Cerium Bromide (CeBr3) scintillator
    \item Bismuth germanium oxide (BGO) scintillator
    \item Low-voltage PMT to reduce the dark rate
\end{itemize}

One can find the discussions of BGO scintillator in \cite{MOSZYNSKI1981403}, while the low-voltage PMT, as well as cryogenic setups, will be directly tested in situ at LANSCE. We further discuss the CeBr3 scintillator design in the following section.
Additionally, we can consider installing Superconducting Nanowire Single Photon Detectors (SNSPDs) to replace the PMTs. SNSPDs are novel quantum sensors with high count rates, sub-picosecond timing jitter, nearly 100 \% quantum efficiency, and zero dark counts \cite{SNSPD2012}.
These sensors can serve as an alternative technology and cross-validate our results.

\subsection{CeBr3 scintillator}

One promising material alternative to plastic scintillator is CeBr3. This new material has six times higher light yield and is five times as dense as the plastic scintillator, leading to a factor of 30 times higher total photon production. It also has a fast scintillation time constant of 20 ns and low intrinsic radioactivity~\cite{Ref_CeBr3}. 
Currently, CeBr3 crystals can be produced with lengths of up to $\sim 20$ cm. However, the cost may be prohibitive, so we consider CeBr3 volumes of 5 cm in length. This shorter 5 cm detector has the same charge sensitivity as the equivalent 1.5 m of plastic scintillator. We can use these shorter detectors to construct a full detector with CeBr3. 

As shown in the simulation section below, the dramatically reduced volume and lower neutron scattering cross-section of CeBr3 allows for a significant reduction in neutron backgrounds, the primary background source at the proposed location. In addition, the small volume of CeBr3 allows for the design of a much thicker neutron shield. These combined effects result in neutron backgrounds being made subdominant compared to dark rate backgrounds for CeBr3 detectors.

\section{mCP Production and Signal}
The main contribution to mCP production at Lujan Center is the decay of neutral mesons, $\pi^{0},\,\eta \to \gamma \chi\overline{\chi}$.
The total number of mCPs is given by 
\begin{equation}
    N_{\chi} \simeq 2 N_{\mathfrak{m}}  \varepsilon_{\chi}^{2}\alpha_{\text{EM}} \text{Br}(\mathfrak{m}\to\gamma\gamma) I^{(3)} \qty(\frac{m_{\chi}^{2}}{m^{2}_{\mathfrak{m}}}).
\end{equation}
Here, $N_{\mathfrak{m}}$ is the number of meson $\mathfrak{m}$ and $I^{(3)}$ is a phase space integral \cite{Kelly:2018brz}.
The meson productions per POT at Lujan Center were previously discussed in \cite{CCM:2021leg} and \cite{Patrick2016, deNiverville:2016rqh}.

\begin{figure}[t]
    \centering
    \includegraphics[width=8.5cm]{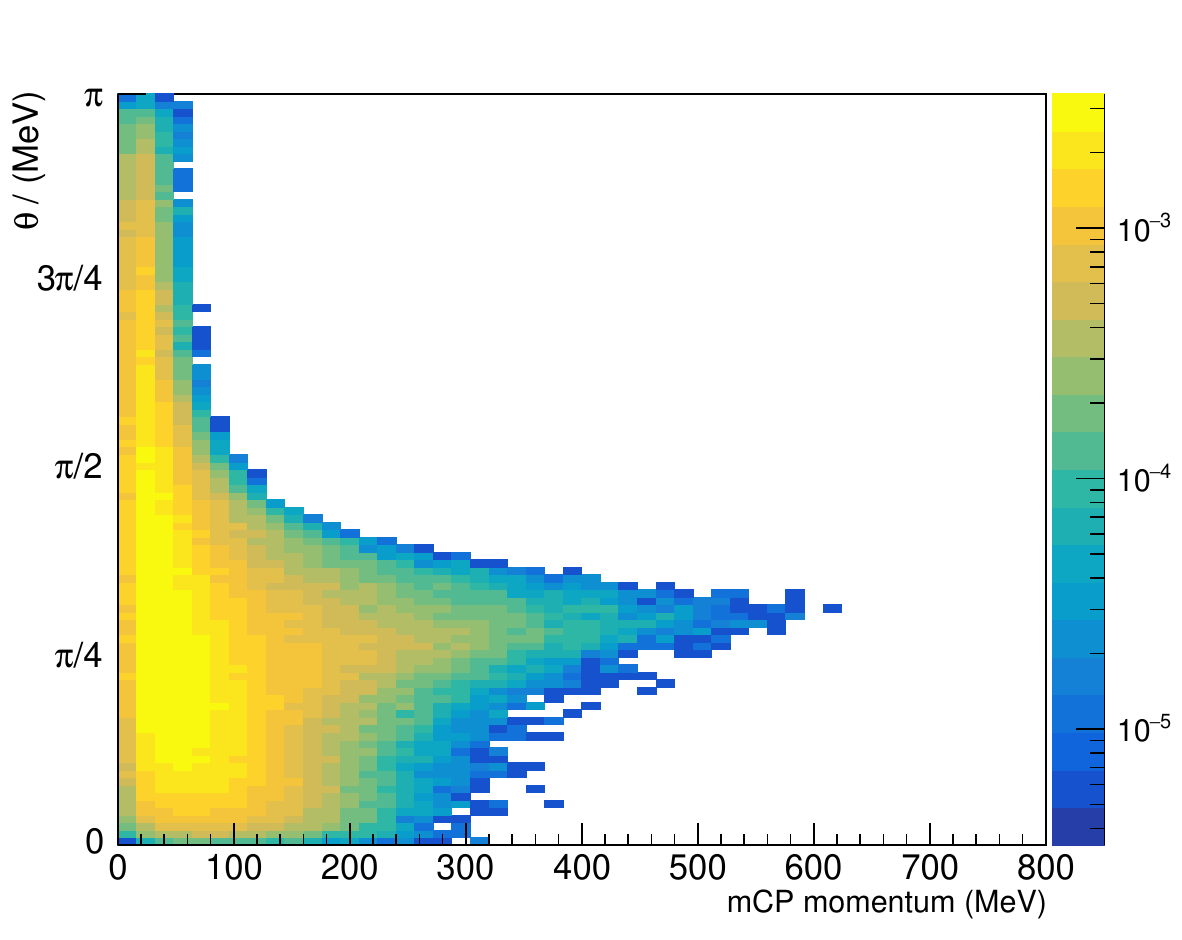}
    \caption{Polar angle and momentum distribution of mCPs with $m_{\chi} = $ 10 MeV produced by $\pi^{0}$ decay.}
    \label{fig:mcp-angleVsmomentum}
\end{figure}

The potential installation sites at Lujan Center are in close proximity to the target but oriented perpendicular to the beam direction. 
To account this, we introduce geometrical acceptance $A_{\text{geo}}$ defined as a fraction of mCPs produced that reach the detector.
We assume a similar momentum distribution with the charged pions for $\pi^{0}$ and $\eta$ \cite{Patrick2016} and generate mesons with the Burman-Smith distribution \cite{Burman:1989ds}.
While the Burman-Smith distribution is currently the available model, further study with dedicated efforts may improve the prediction of the meson spectrum.
This exceeds the scope of this proposal and is left for future study.
The decay $\mathfrak{m} \to \gamma\chi\overline{\chi}$ is then simulated according to  
\begin{align}
&\frac{d\text{Br}(\mathfrak{m} \!\to\! \gamma\chi\overline{\chi})}{ds \, d\!\cos\theta}
 = \frac{\varepsilon_{\chi}^{2} \alpha_{\text{EM}}}{4 \pi s} \qty\bigg(1\!-\!\frac{s}{m_{\mathfrak{m}}^2})^3 \qty\bigg(1-\frac{4m_\chi^2}{s})^{\frac12} \nonumber \\
 & \quad \times \qty\bigg[2-\qty\bigg(1-\frac{4m_\chi^2}{s} ) \sin^2 \theta]  \text{Br}(\mathfrak{m} \!\to \!\gamma\gamma),
\end{align}
where $s = (p_\chi + p_{\bar\chi})^2$ is the invariant mass of the off-shell photon $V^{*}$ and $\theta$ is the polar angle of $\chi$ momentum measured in the rest frame of $V^{*}$ with its $z$-axis aligned to the boost direction of $V^{*}$~\cite{Jodlowski:2019ycu}.
Going into the lab frame, we obtain the angular distribution of mCPs. FIG \ref{fig:mcp-angleVsmomentum} illustrates the angular distribution of mCP with mass of 10 MeV produced by $\pi^{0}$ decay.
The $A_{\text{geo}}$ is estimated for a range of mCP mass $m_{\chi}$ and the detector located at various distances from the target, whose cross-sectional area is $5 \text{ cm} \times 5 \text{ cm} \times \text{100 bars} =  0.25 \; \text{m}^{2}$.
We find a probability of $\mathcal{O}(10^{-4})$ for mCPs to reach the detector at ER1 and $\mathcal{O}(10^{-5})$ at ER2. 

Any mCPs entering the detector will pass through the all $n$ layers, producing photons that can be detected by PMTs.
The probability of observing a coincidence signal of $n$ layers follows the Poisson distribution $ P = (1-\exp[-\varepsilon_{\chi}^{2} N_{\text{PE}}])^{n}$, 
where $\varepsilon_{\chi}$ is the fractional charge of the mCP and $N_{\text{PE}}$ is the number of photoelectrons for $\epsilon_{\chi} = 1$.
The $N_{\text{PE}}$ is proportional to the number of photons generated in a scintillator reaching the photocathode of a PMT, the quantum efficiency of the PMT, and the mean energy loss of a charged particle in a scintillator bar \cite{pdg2022}.
The factor $\varepsilon_{\chi}^{2}$ has been introduced since the energy deposition of a charged particle within the scintillator is proportional to $Q^{2}$.
A {\textsc{Geant4}\xspace}~\cite{GEANT4} simulation has been made to find  $N_{\text{PE}} = 2.5\times 10^{5}$ for a  $5\,\times\,5\,\times150\; \text{cm}^{3}$ Eljen-200 (EJ-200) scintillator wrapped in $97 \; \%$ reflective white Tyvek paper and coupled to a Hamamatsu R7725 PMT.
The expected total number of signal events $s$ is given by $s = PA_{\text{geo}}N_{\chi}$.

\section{Background Determination}

In this section, we discuss the simulations, measurements, and shielding of the background. The dominant physical background sources are neutrons from the beam, along with beam-induced and ambient photons. The beam can irradiate materials in the proposed detector locations which can generate ambient neutron and gamma backgrounds. Borated polyethylene and water are effective in reducing neutron backgrounds, and lead shielding is effective in reducing gamma backgrounds. A mix of shielding materials can, therefore, be used to reduce these beam-induced backgrounds.

\subsection{Simulations}
\begin{figure}[t]
    \centering
    \includegraphics[width=9cm]{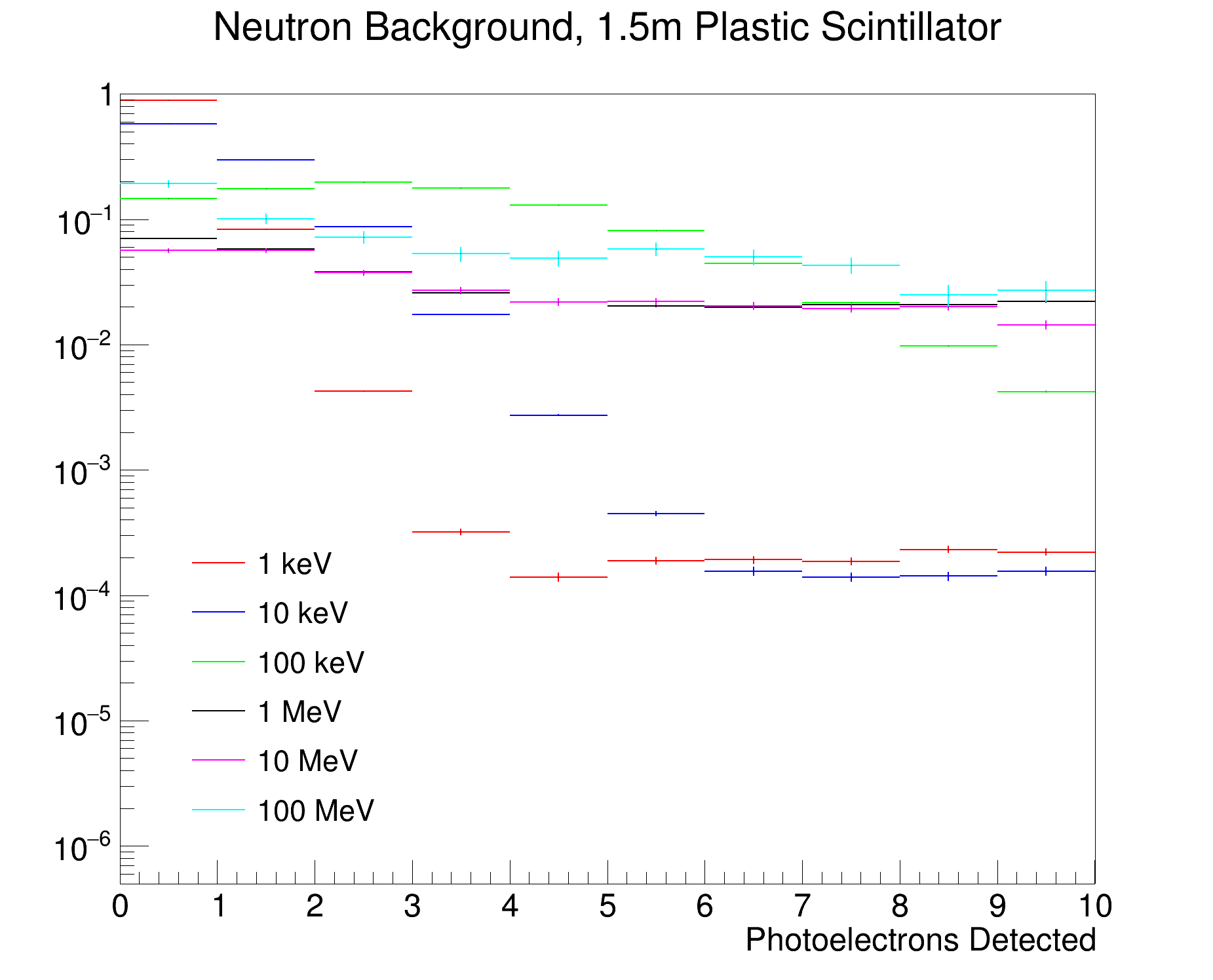}
    \caption{Neutrons of a range of energies produce signal-like deposits in the plastic scintillator, owing to the large neutron scattering cross-section of hydrogen.}
    \label{fig:plastic-neutron}
\end{figure}

\begin{figure}[t]
    \centering
    \includegraphics[width=9cm]{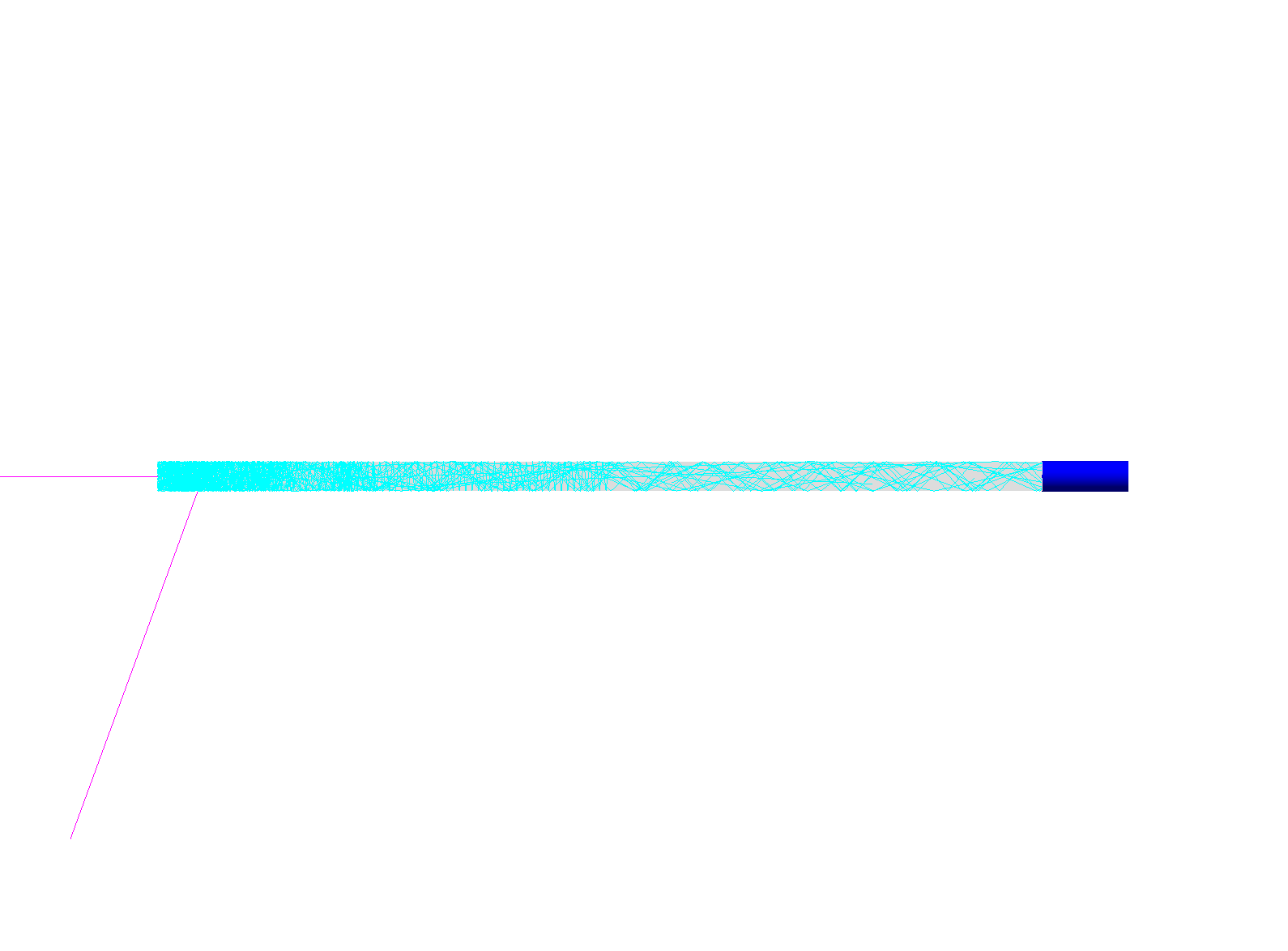}
    \includegraphics[width=9cm]{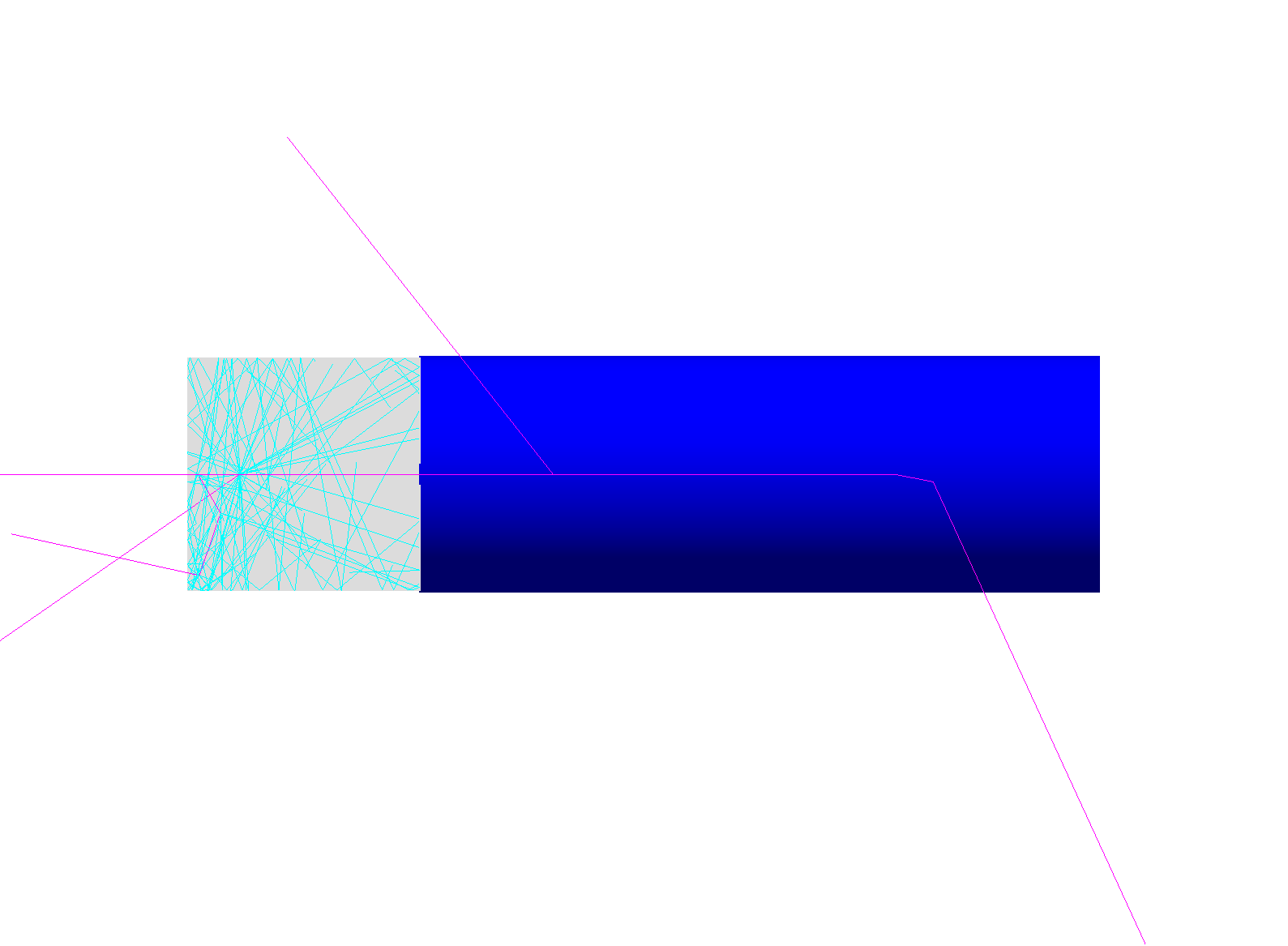}
    \caption{Neutrons (purple) of energy 10~keV scatter softly in scintillator, producing a small, signal-like number of scintillation photons (cyan) more frequently than any other neutron energy. These scintillation photons are collected in a PMT (blue).}
    \label{fig:sim-neutron}
\end{figure}

Given the significant neutron flux at the proposed detector location, understanding backgrounds arising from soft neutron scatters in scintillator is critical to detector design and sensitivity estimates. A {\textsc{Geant4}\xspace} simulation was used to model the impacts of these soft neutron scatters in a 150 cm plastic scintillator bar and in a 5~cm CeBr3 detector, each with a 5$\times$5~$\text{cm}^{2}$ cross-section, illustrated in FIG~\ref{fig:sim-neutron}. A geometry with a neutron and photon shield consisting of 10~cm each of 5 \% borated poly, water, and lead was studied in addition to an unshielded geometry.
In the case of plastic scintillator, FIG~\ref{fig:plastic-neutron} illustrates the resulting photoelectron distribution for a range of input neutron energies. Signal-like energy deposits are in the 1-3 photoelectron range. At very low neutron energies ($<10$ keV), too little energy is deposited to consistently generate a detectable signal. At very high energies ($>10$ MeV), soft, signal-like neutron scatters are increasingly rare as the energy deposits follow a flat distribution in the number of photoelectrons detected. Therefore, the intermediate energy region between 10 keV to 10 MeV is of most interest for shielding designs and background estimation.

\begin{figure}[t]
    \centering
    \includegraphics[width=9cm]{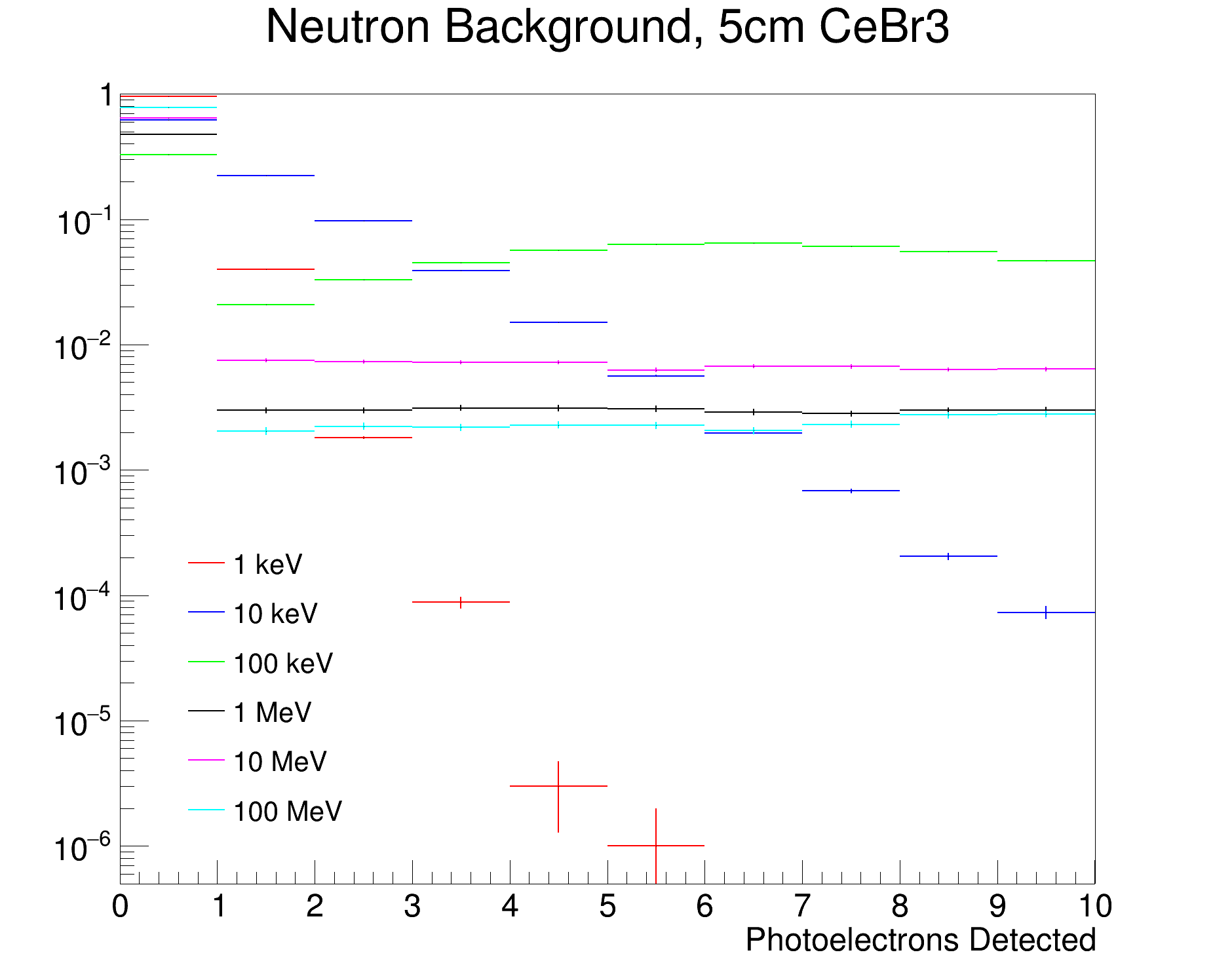}
    \caption{The small volume of CeBr3, in addition to decreased neutron capture cross-sections, leads to a significant reduction in backgrounds.}
    \label{fig:cebr-neutron}
\end{figure}

Compared to a plastic scintillator, a CeBr3 detector has significantly reduced neutron backgrounds, as shown in FIG~\ref{fig:cebr-neutron}. The significant decrease in material along with a reduction in neutron scattering cross-section for cerium and bromine compared to hydrogen results in this effect~\cite{NeutronNews}. 
The relative incidence of signal-like soft neutron scatters behaves similarly to plastic, but with an overall decrease in rate.

\begin{figure}[t]
    \centering
    \includegraphics[width=9cm]{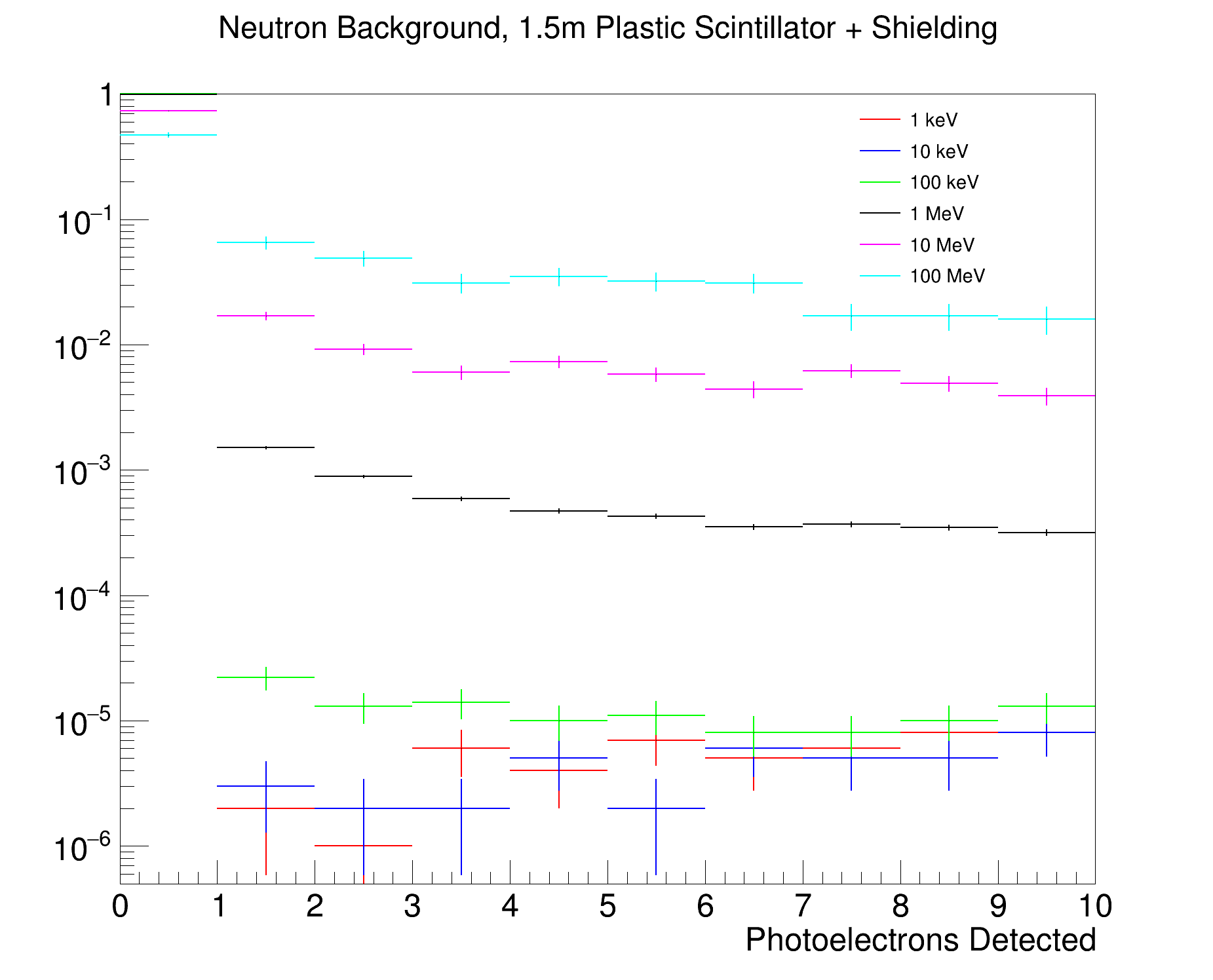}
    \includegraphics[width=9cm]{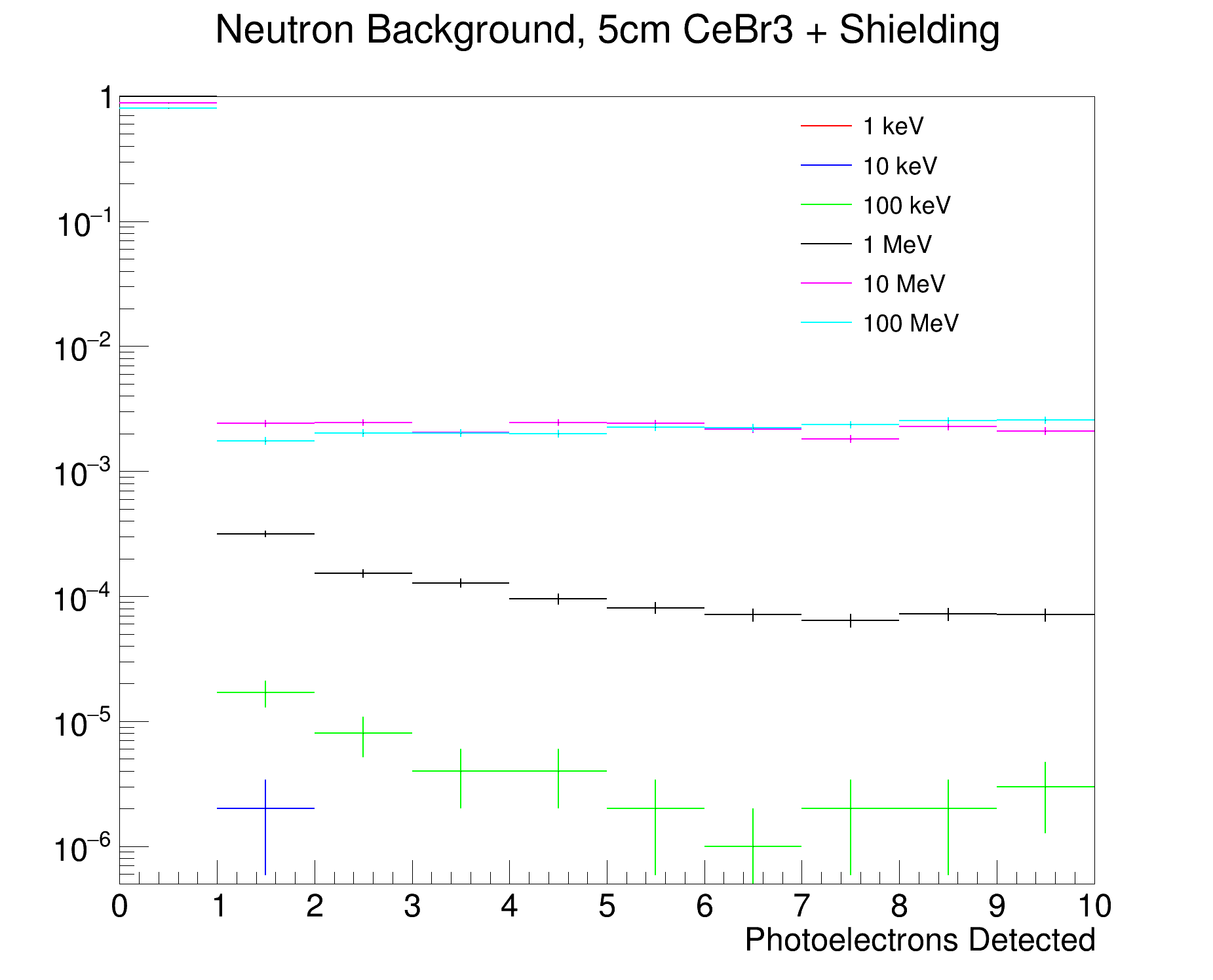}
    \caption{Added neutron shielding captures low energy neutrons, leaving only the broad, flat photoelectron distributions from neutrons at energies of 1~MeV and greater.}
    \label{fig:shield-neutron}
\end{figure}

When applying the neutron shielding scheme, neutrons of energy below 1~MeV are captured at high probability, leading to a reduction in neutron-induced backgrounds as illustrated in FIG~\ref{fig:shield-neutron}. 
As discussed below, in situ measurements conducted with a plastic scintillator and a shield consisting of 5~cm of Pb and 5~cm of borated polyethylene suggest that 30~cm of neutron shielding material is required to suppress neutron backgrounds to a level under dark rate.

\subsection{In situ measurements}
\begin{figure}[t]
\centering
\includegraphics[width=9.0cm]{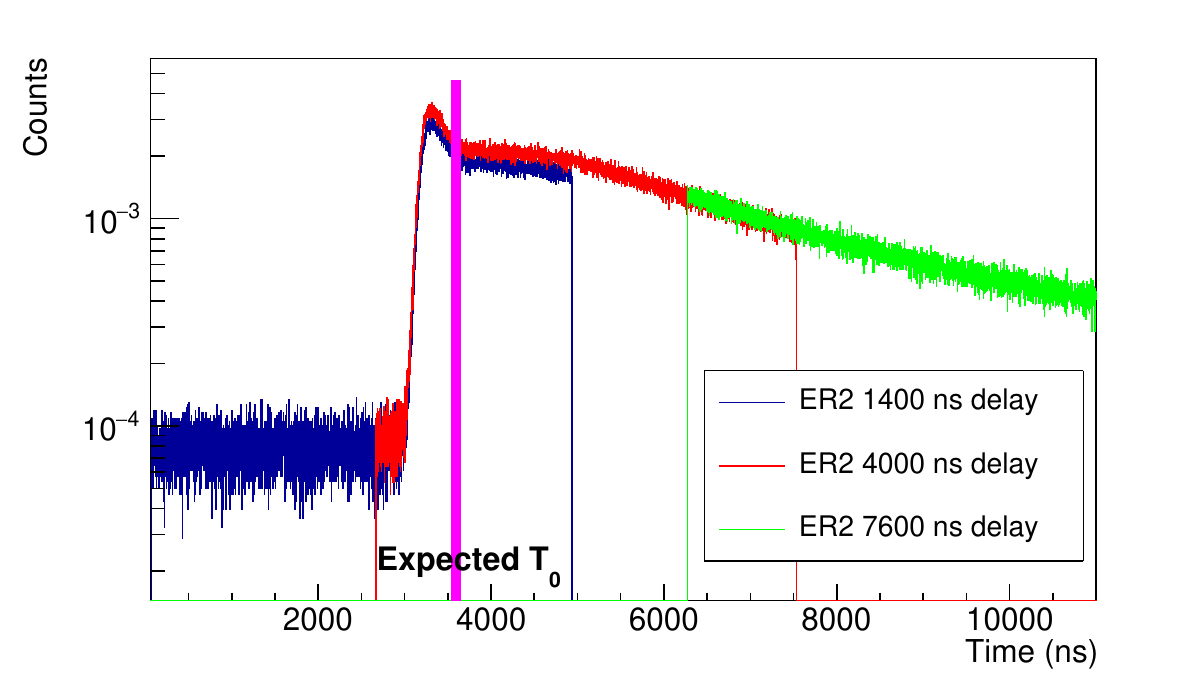}
\caption{Pulse counts per event over time at ER2 generated by concatenating 3 runs with different trigger delays. The pink line indicates the beam trigger time, which is about 500 ns behind the actual rise time of the pulse counts. The number of pulses increases by an order of magnitude after the beam incident.
}
\label{fig:pulse_time}
\end{figure}

In situ measurements have been performed at the potential detector site at both ER1 and ER2 to estimate the background rate and determine the optimal shielding configurations.
A custom-designed readout system cascading 4 input channels of a domino ring sampler-4 (DRS4) chip \cite{DRS4} to obtain a 5 \textmu s acquisition window was used for data acquisition.
As mentioned in section~\ref{sec:facility}, the LANSCE facility provides the T0 signals for each proton impact on the target by which the readout system was triggered.
FIG~\ref{fig:pulse_time} shows an example of this over 11 \textmu s range, generated by concatenating 3 runs with different trigger delays.
We used EJ-200 plastic scintillators measuring 5~cm $\times$ 5~cm $\times$ 80~cm, coupled with Hamamatsu R7725 PMTs (hereafter referred to as modules) operating at 1300 V.

\begin{figure}[t]
    \centering
    \includegraphics[width=8.7cm]{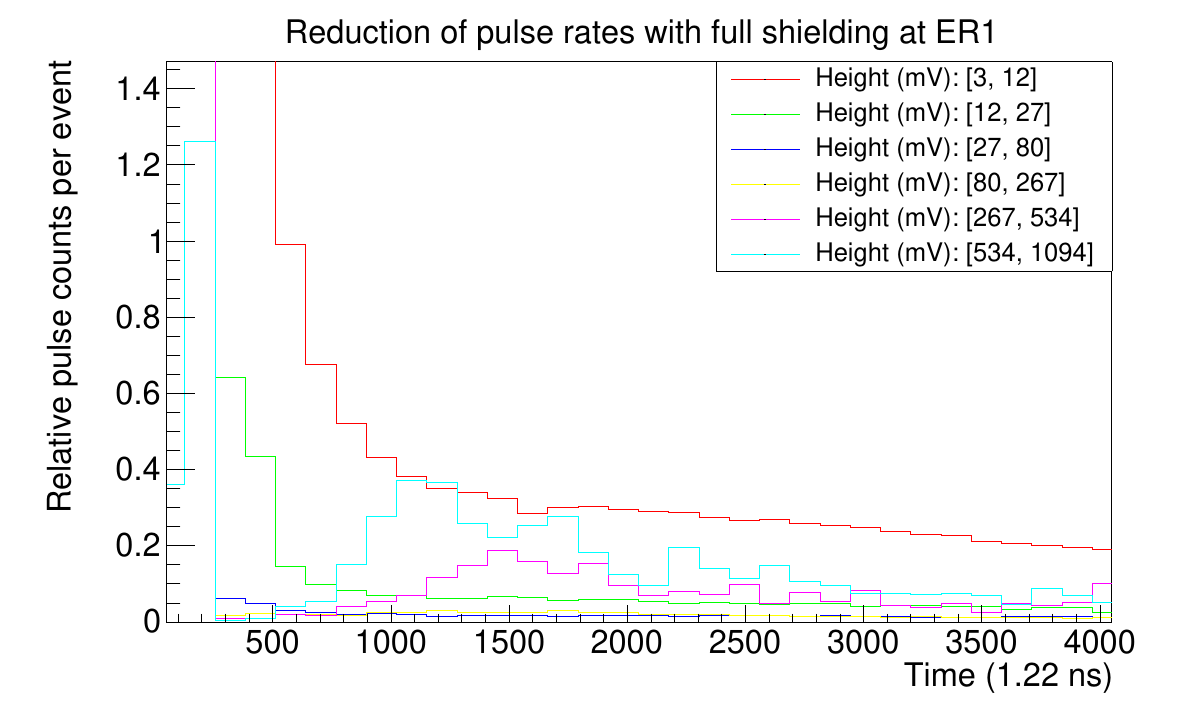}
    \includegraphics[width=8.7cm]{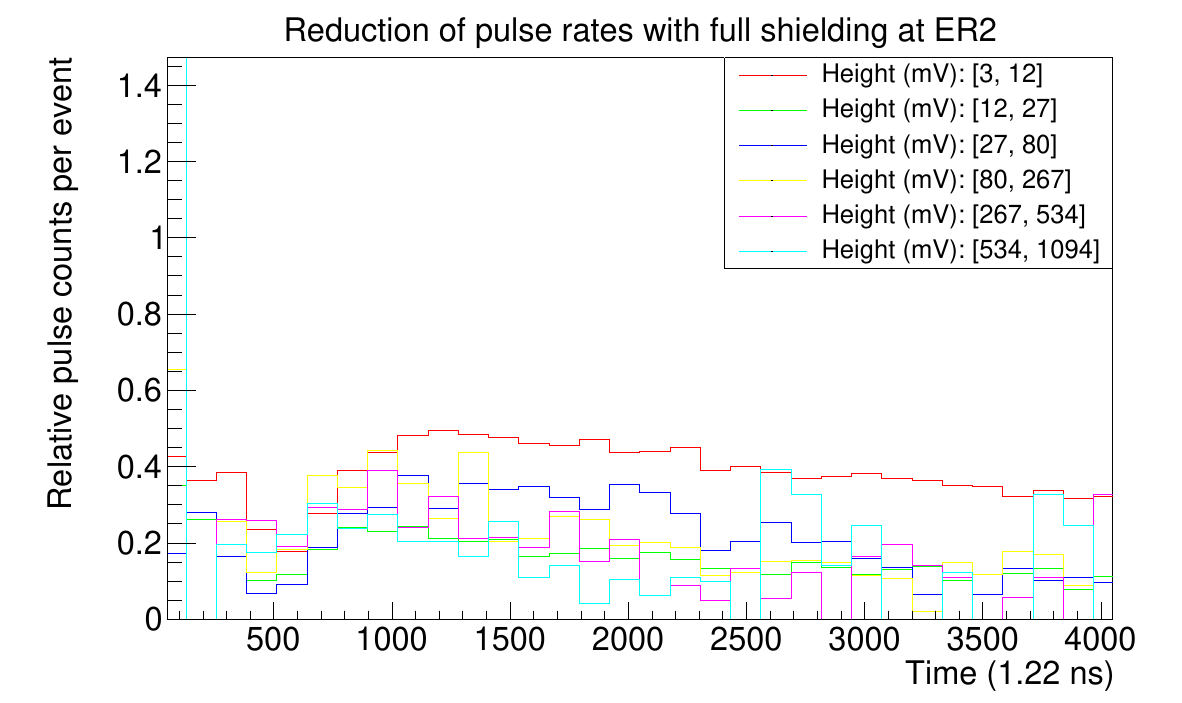}
    \caption{Relative number of pulses over time in a certain pulse height range with full shielding compared to that of bare setup at ER1 (upper panel) and ER2 (below panel). The beam trigger is at 800 TDC, and the expected arrival time of particles from the beam is at 450 TDC.
}
    \label{fig:shielding_effect}
\end{figure}

{\bf Shielding effect:} The shielding effects of $5$~cm thick polyethylene, $5$~cm thick lead, and their combination (full shielding) were studied by comparing the response of the same module with and without the shielding.
The plastic shielding alone resulted in a reduction of pulse counts by 70~\% at ER1 and 60~\% at ER2. 
When combined with the lead shielding, there was a further reduction of pulse counts by 80~\% at ER1. 
Lead shielding alone exhibited no discernible effect at ER1 but resulted in a 60~\% reduction in pulse counts at ER2, suggesting higher radiation-induced backgrounds at ER2. 
Since other experiments and facility structures are situated between the target and the detector module at ER2, it is possible that these factors contribute to the scattering of energetic particles, producing more gamma rays and other secondary particles compared to ER1.
FIG~\ref{fig:shielding_effect} illustrates the effect of shielding on a specific pulse height range by counting the number of pulses measured with shielding within a designated pulse height range, relative to those obtained without shielding.
The pulse height corresponding to a single photoelectron signal of the PMTs used, which is the mCP signature, is approximately 8 mV, and a 70 \% reduction rate in pulse counts due to full shielding is observed within this range.
The detail of the background reduction of the full shielding would depend on the energy and angular distributions of the neutrons, as well as the secondary scattering effects. 
However, we expect that the shielding with a similar chemical combination of the tested shielding material with a thickness of $> 30$ cm would reduce neutron backgrounds to a level comparable to the backgrounds from dark counts. Further, in situ studies will be performed to optimize the final design of this thicker shield.

{\bf Coincidence rate:} Two bare modules placed in a row were used to estimate the coincidence event rate by detecting subsequent pulses from both modules with 20~ns time window. 
We considered pulses occurring within 200 ns after the beam trigger for both modules to avoid neutron flux \cite{CCM:2021leg}.
A pulse height cut of $<$ 16~mV was applied to veto large pulses due to beam-related particles and cosmic events.
For 3 hours of running, the number of coincidence events per 1,000 collisions is $7.7\pm 0.2$ at ER1 and $1.2 \pm 0.1$ at ER2.
These rates imply that, over 3 years of operation, an unshielded neutron background rate is of order $10^8$ at ER1 ($10^7$ at ER2), compared to an expected dark count rate of $10^2$.
We assume that the two layers are spatially separated and have uncorrelated neutron events for simplicity. Considering 60-70 \% reduction in pulse rates offered by 5 cm polyethylene shielding for each layer, we estimate that a 30 cm neutron shield would mitigate the neutron background in both sites to a level comparable to the dark current-induced background.
Additional reductions in neutron rates are expected for scintillator volumes nearer to the center of each layer, as the scintillator itself in the outer regions of each layer offers self-shielding effects.

\subsection{Dark current}
Another background source is the random coincidence signal event due to the dark current of PMTs.
The random coincidence rate $R$ of $n$ layers is given by $v_b^{n}\tau^{n-1}$, where $v_{b}$  is the dark count rate, and $\tau$ is the coincidence time window.
Since the total number of random coincidence signals is proportional to the data tracking time, the beam timing information can be incorporated to significantly reduce the dark current backgrounds.
For the proton beam with $280$ ns bunch-width, we set the data acquisition window as $500$ ns after the beam trigger signal to capture most of the mCP events.
Once we know the live time, the total number of random coincidence signals is given by $R \times \text{total live time} \times \text{number of modules per layer}$.

\subsection{Cosmic background}

Given the very short trigger live time, the cosmic ray-induced background is subdominant, as discussed in detail in the SUBMET experiment with a similar environment \cite{SUBMET-cosmic2023}.
For the proposed 10$\times$10 detector with 2 layers, nearby cosmic muon flux has a rate of approximately 1 kHz. Rarely, these muons generate gammas from secondary EM showers which deposit energy in the detector. According to estimates made for the SUBMET detector, a similar 2-layer mCP detector, these cosmic backgrounds are expected to be subdominant \cite{Kim:2021eix}. Added shielding designed to reduce beam-induced photon and neutron backgrounds offers additional protection against cosmic showers. These backgrounds will be measured during beam-off periods to ensure the contributions are small.

\subsection{Summary of background reduction strategy}

To significantly reduce the beam-related backgrounds, we require at least 30 cm shielding with polyethylene and/or water, combined with the incorporation of the precise arrival time of mCPs and neutrons, and elaborated pulse finding with characterization of neutron signals and afterpulses.

In addition, the compact size of a CeBr3 scintillation detector, with significantly reduced neutron interaction length and cross-section, would result in reduced neutron backgrounds and allow larger space for shielding materials in comparison to plastic scintillators.

\section{Sensitivity Projections}

\begin{figure}[t]
    \centering
    \includegraphics[width=9.3cm]{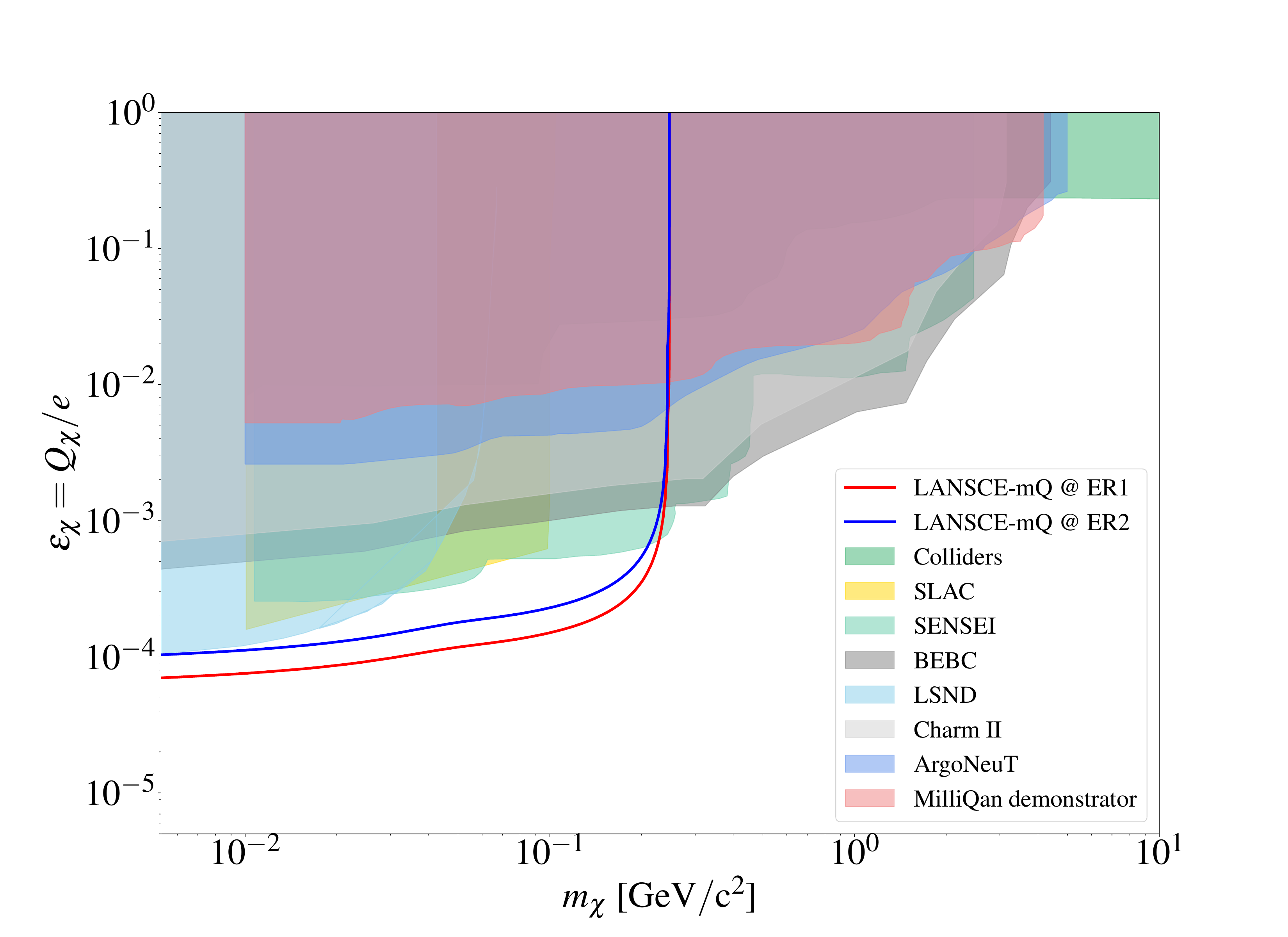}
    \caption{The sensitivity projection of the 95\% CL limit curves of the nominal design of LANSCE-mQ for 3 years of operation accumulating $5.9\times 10^{22}$ POT at ER1 (red curve) and ER2 (blue curve).}
    \label{fig:sensitivity_final}
\end{figure}

In FIG~\ref{fig:sensitivity_final}, we show the $95$ \% CL exclusion limit of the LANSCE-mQ detector with nominal setup over 3 years of operation at ER1 and ER2. 

Considering the background sources discussed in the previous section, we expect to use sufficient shielding (around 30~cm of borated polyethylene or water) to reduce the rate of neutron backgrounds to be the same order as dark current-induced backgrounds.
In this plot, we assume a typical dark count rate of $\sim 200$ Hz~\cite{Meyer:2008qb,Foroughi-Abari:2020qar} for a R7725 PMT, resulting in dark current-induced coincidence events over $3$ years of live time of $150$. We then assume twice the number of neutron backgrounds and $50$ events from cosmic and environmental radiation sources, to predict a total background  over $3$ years of operation to be $500$ events.
These numbers are determined based on our knowledge from dedicated simulation, in situ background measurements, and estimations of the shielding effects. Further in situ measurements can help determine the systematic errors for the full experiment.

One interesting feature of the sensitivity projection is that the curve is smooth when the mCP becomes heavier and goes from pion-dominant production to $\eta$-dominant production, compared to other accelerator searches depending on meson productions. 
The reason for this is that $\eta$ mesons produced by the 800 MeV beam are nearly at rest, leading to more isotropic mCP production when they decay.
Given the unique Lujan beam and detector configuration in which the detectors are located perpendicular to the beam direction, the isotropic production enhances the sensitivity compared to the pion production, resulting in a smooth transition.

\section{Discussions}

Now that we have established that LANSCE-mQ provides a leading search for sub-GeV mCPs, we discuss several related experiments and sites that can further the mCP searches with similar productions or search techniques.
Most notably, at CCM~\cite{CCM:2021leg}, one experiences a similar level of mCP flux as discussed in this paper, and it has a larger detector volume. Although not a dedicated mCP search, one can study electron signatures from mCP scattering with this experiment.

{\bf Alternative sites for dedicated mCP searches:} The studies in this paper motivate explorations of dedicated millicharged detectors at other neutron facilities, including n\_TOF at CERN~\cite{nTOF} and the future European Spallation Source~\cite{Abele_2023}. Other promising sites for dedicated mCP searches include the {\it backroom} behind the 10 m iron block of SpinQuest/DarkQuest at Fermilab, where LongQuest~\cite{Tsai:2019buq} was proposed. 
Millicharge searches at SHiP~\cite{Magill:2018tbb} with dedicated detectors can also provide excellent sensitivity for mCPs.
The drastically different background sources, as well as beam structures, require dedicated simulations and measurements, so we leave them for future studies.

\section{Acknowledgements}

We thank Justin Evans, Teppei Katori, William Louis, and Richard Van de Water for useful discussions. 

This research is partially supported by LANL's Laboratory Directed Research and
Development (LDRD) program. YDT thanks the generous support from the LANL Director's Fellowship.
This research is partially supported by the U.S. National Science Foundation (NSF) Theoretical Physics Program, Grant No.~PHY-1915005. This research was supported in part by grant NSF PHY-2309135 to the Kavli Institute for Theoretical Physics (KITP).
This work was partially performed at the Aspen Center for Physics, supported by National Science Foundation grant No.~PHY-2210452. This research was partly supported by the National Science Foundation under Grant No.~NSF PHY-1748958. This document was partially prepared using the resources of the Fermi National Accelerator Laboratory (Fermilab), a U.S. Department of Energy, Office of Science, HEP User Facility. Fermilab is managed by Fermi Research Alliance, LLC (FRA), acting under Contract No. DE-AC02-07CH11359. This work was partially supported by the National Research Foundation of Korea (NRF) grant funded by the Korean government (MSIT) (No. 2021R1A4A102297713).

\bibliography{references}
\end{document}